\documentclass[prb,twocolumn,twocolumn,superscriptaddress,showpacs]{revtex4}
\usepackage{amsmath, amsfonts,amssymb, amsthm,bbm}
\usepackage[utf8]{inputenc}
\usepackage{natbib} 
\usepackage{graphicx}
\usepackage{color}

\newcommand{\im}{\operatorname{Im}} % Imaginary Part of complex numbers
\newcommand{\re}{\operatorname{Re}} % Real Part of complex number 
\newcommand{\unibasengl}{Department of Physics, University of Basel, Klingelbergstr. 82, 4056 Basel, Switzerland}
\newcommand{\kBT}{k_\text{B}\! T}
\newcommand{\mb}[1]{\text{\boldmath ${#1}$}}
\def\ket#1{\mathinner{|{#1}\rangle}}% Vector
\def\bra#1{\mathinner\langle{#1}|}% Dual Vector
\def\fbkt#1#2#3{\mathinner{\langle{#1}\,|\,#2\,|\,#3\rangle}} % Full operator scalar-product
\newcommand{\abs}[1]{\left\lvert#1\right\rvert}% modulus
\newcommand{\abss}[1]{\lvert#1\rvert}% small modulus
\newcommand{\sgn}{\operatorname{sgn}} % signum function
\newcommand{\Pint}{\mathcal{P}\negthickspace\negthickspace\int} % principal value integral
\newcommand{\D}{{d}}% Integration
\newcommand{\Id}{\textrm{Id}}% Identity
\newcommand{\Tr}{\operatorname{Tr}} % Trace
\newcommand{\erw}[1]{\langle {#1} \rangle}% Expectationvalue
\newcommand{\proj}[1]{\ket{#1}\bra{\!#1}}% Projecting a single variable
\begin{document}
%% Frontmatter
\title{Quantum transport through single-molecule junctions with orbital degeneracies}
\author{Maximilian G.~Schultz}\email{maximilian.schultz@unibas.ch}
\affiliation{\unibasengl}
\date{\today}
\begin{abstract} 

  We consider electronic transport through a single-molecule junction where the molecule has a degenerate spectrum. Unlike previous transport models,
  and theories a rate-equations description is no longer possible, and the quantum coherences between degenerate states have to be taken into account.
  We present the derivation and application of a master equation that describes the system in the weak-coupling limit and give an in-depth discussion
  of the parameter regimes and the new phenomena due to coherent on-site dynamics.

\end{abstract} 
%% 81.07.Nb	Molecular Nanostructure
%% 05.60.Gg	Quantum Transport
%% 73.23.Hk	Coulomb blockade and single-electron tunneling
%% 85.65.+h	Molecular electronic device
\maketitle
%% Body

\section{Introduction} %% <<<

The field of molecular electronics has greatly benefited from the possibility of formulating transport problems in the weak-coupling regime as rate
equations for sequential tunneling processes.\cite{Braig03,Boese01,Huettel09,Koch05,Recker08,Romeike06,Wege05,Sapmaz06,Leturcq08} Given the spectrum of
the molecule and the tunneling-matrix elements, the different features of the steady-state current--voltage characteristics can immediately be mapped
onto the energetic availability or non-availability of certain jump processes.  Rate equations are the Markovian kinetic equations for the
lowest-order expansion of the von Neumann equation with respect to the tunneling Hamiltonian while neglecting any off-diagonal elements of the reduced
density matrix. Such an approximation is well justified for non-degenerate systems, when the differences of the molecular eigenenergies are much
larger than any tunneling-induced level shift or broadening. As molecules often feature geometric and thus orbital symmetries, the system Hamiltonian
of a single-molecule junction shows degenerate levels.  Due to the quantum-mechanical nature of the tunnel junction, the rate-equation description is
in general inadequate for these systems. 

The problem of electronic transport through quantum nanostructures with degenerate levels has already been given some attention in different parts of
the literature. The comprehensive review of Markovian master equations by Timm \cite{Timm08} shows the equivalence of different methods and approaches
used to derive master equations for weak-coupling problems. In Refs.~\onlinecite{Braun04} and \onlinecite{Braig05}, the coupling of a spin-degenerate
quantum dot to ferromagnetic leads causes coherent dynamics described by the full master equation which significantly differ from the spectroscopic picture
found in rate-equation treatments.  The problem of using the rate-equation formalism for molecules with orbital symmetries has already been addressed
in our previous study on Jahn--Teller molecules.\cite{Schultz08} An implementation of the full master-equation formalism for a genuine
molecular-electronics problem is discussed in Refs.~\onlinecite{Begemann08} and \onlinecite{Darau08}, however, only for a very special and rather
complex type of molecule. 

The dynamics of a master equation\footnote{We shall use the term ``master equation'' as a synonym for master equation for the full reduced density
matrix, in order to contrast this equation to the rate equation, which only concerns the density matrix's diagonal elements.} is, in contrast to a
rate equation, much less intuitive as the tunneling electrons are allowed to jump into and out of linear superpositions of the degenerate molecular
states. The coupling to the continuum of states in the electronic reservoirs via virtual transitions generates an intrinsic dynamics on the molecule
that is not related to real changes of the number of electrons in the system. The image of the dynamics as a succession of well-defined quantum jumps
between the leads and the molecule, which renders the rate equation so simple in its use and tempting for application, is declared void by the quantum
mechanical nature of the degenerate system. 

The basic phenomenology of quantum transport through nanostructures with orbital degeneracies in the absence of vibrations and in particular the
proper derivation of a Markovian master equation for the treatment of near-degeneracies has already been investigated by us in Ref.~\onlinecite{Schultz08c}.
There we establish the ``decoupling paradigm'', which states that for generic tunnel amplitudes, there is always a basis of the molecular Hilbert
space where one of the (near-)degenerate levels is decoupled from the drain electrode. Below the double-charging threshold, this level is rendered a
dark state in which charge is accumulated, and electronic transport across the nanostructure is strongly inhibited. This coherent current-blockade is
only partially lifted due to the tunneling-induced renormalization of the isolated structure's levels, as by transitions via virtual intermediate
states in the source electrode, the electron is moved out of the dark state and allowed to tunnel to the drain electrode. A particularly appealing
interpretation of this dynamics uses the picture of a pseudo-spin and the tunneling-induced renormalization as a pseudomagnetic field acting on that
pseudo-spin.\cite{Braun04}

The purpose of this article is the extension and discussion of the model to single-molecule devices, mainly the relation of the physics caused by the
orbital degeneracies to the vibronic dynamics of the molecular cage. After having defined the most general model of a linear electron--phonon
coupling, we shall show that for the study of degenerate systems this can be reduced to two genuinely different types of molecular models, which
significantly differ in their transport properties. One of them, which we shall term ``Anderson--Holstein model'', will simply show a superposition of
a vibronic sideband structure and the already known coherent current blockade generic for degenerate electronic systems. The ratio of charging energy
to vibronic energy will be shown to characterize the steady-state current--voltage profile: for large charging energy, the vibronic sidebands will
appear as peaks instead of steps and thus render the appearance of negative differential conductance a generic property of single-molecule junctions
with degenerate orbitals. The other, a Jahn--Teller active model, will show a strong dependence of the electronic transport properties on the coupling
to the leads.

In the last section, we shall extend the basic model incorporate elements that are related to possible experimental issues including the modifications
of the transport properties due to slight breaking of the orbital degeneracy, general linear electron--phonon coupling, and the presence of many modes
in the electronic reservoirs. 

Not surprisingly, our results will reproduce certain effects and features, which have already been reported by several
groups;\cite{Braun04,Darau08,Begemann08,Schaller09} what we, however, do want to show is that by using a bottom-up approach and adding complexity to the models
in several steps, we succeed in tracing the fundamental and generic physics back to the intrinsic properties of the master equation and are thus in a
much better position to actually apply the theory to experiments. 

%% >>>

\section{General Properties of the Hamiltonian} %% <<<

The model of a single-molecule junction, as we consider it in this article, consists of three parts: the single molecule itself, $H_\text{mol}$, the
source and the drain electrodes through which electrons are injected and extracted, $H_\text{leads}$, and a tunnel-coupling between the two,
$H_\text{T}$. The electrodes, $\alpha$ being an index for left and right, are modeled as spinless, non-interacting Fermi gases, $H_\text{leads} =
\sum_{\mb{k}\alpha} \varepsilon_\mb{k} c^\dagger_{\mb{k}\alpha} c_{\mb{k}\alpha}$, in the wide-band limit, that is with constant density-of-states
$\nu_0$.  The molecule, in order to distinguish it from quantum dot systems, is a discrete electronic system coupled to a single vibrational
mode.\cite{Braig03,Mitra04,Koch05} The electronic part in our model is rather simple. We consider two degenerate levels that can be detuned from their
energy $\varepsilon_0$ by the gate-voltage of the molecular junction, a Coulomb interaction of strength $U$ between the two, but no intra-molecular
tunneling. A small energy difference between the two levels of the order of the tunneling-induced level shift can easily be included in the model and
the thus resulting derivation of the master equation by application of the singular-coupling limit,\cite{Schultz08c} modifies the equation only
marginally; we shall return to this later in section~\ref{sec:SCL}. The molecular vibrations are modeled as a harmonic oscillator of frequency
$\omega$ being coupled to the electronic degrees of freedom by $H_\text{el--ph}$ specified in the next section. The Hamiltonian of the molecule is
thus
\begin{equation}
H_\text{mol} = (\varepsilon_0 + eV_\text{g}) (n_\uparrow+n_\downarrow) + U n_\uparrow n_\downarrow + \hbar\omega b^\dagger b + H_\text{el--ph}.
\end{equation}
The two degenerate electronic levels are, in analogy to the notation of spin, labeled $\ket{\uparrow}$ and $\ket{\downarrow}$. In case we have to sum
over the different levels, we switch to denoting the levels by $\ket{\sigma}$ and its opposite $\ket{\bar{\sigma}}$.  The tunneling part, where from
the beginning, we assume the amplitudes $t_{\alpha\sigma}$ to be independent of the electrons' wave vector, is
\begin{equation}
H_\text{T} = \sum_{\mb{k}\alpha\sigma} t_{\alpha \sigma} c^\dagger_{\mb{k}\alpha}d_\sigma + \text{h.c.}
\end{equation}
For notational convenience, we define a coupling tupel $\mb{\Gamma} := \left(\Gamma_{\text{L}\uparrow}, \Gamma_{\text{L}\downarrow},
\Gamma_{\text{R}\uparrow}, \Gamma_{\text{R}\downarrow}\right)$ being derived from the respective Golden-Rule expressions $\Gamma_{\alpha\sigma} :=
\frac{2\pi}{\hbar}\nu_0 \abss{t_{\alpha\sigma}}^2$.  

\subsection{Electron--Phonon coupling and Degeneracies} %% <<<

The general form of a linear coupling of the oscillator's coordinate to the charge number of a degenerate two-orbital molecule is
\begin{equation}
H_\text{el--ph} = \hbar\omega\sum_{\sigma,\tau = \uparrow, \downarrow}\lambda_{\sigma\tau} d^\dagger_\sigma d_\tau (b^\dagger + b).
\end{equation}
When the expression for $H_\text{el--ph}$ is diagonalized in the space of degenerate electronic levels, which is necessary for the derivation of the
master equation, the only non-trivial couplings are those to the excess charge and the charge difference, respectively,
\begin{equation}
H_\text{el--ph} = \lambda\hbar(b^\dagger + b)\bigl(\lambda_0(n_\uparrow + n_\downarrow) + \lambda_z (n_\uparrow - n_\downarrow) \bigr).
\end{equation}
The canonical transformation generally used to eliminate the electron--phonon coupling---the polaron
transformation\cite{Mahan00}---induces a renormalization of the electronic eigenenergies---the polaron shift---which is proportional to
the square of the electron--phonon coupling strength. The energy of the state $\ket{\uparrow}$ will thus be renormalized by $(\lambda_0 +
\lambda_z)^2\hbar\omega$ and the one of $\ket{\downarrow}$ by $(\lambda_0 - \lambda_z)^2\hbar\omega$, accordingly. The Hamiltonian in the polaron
picture is, however, the starting point for the perturbative analysis of the von Neumann equation, and the question, whether we deal with a
degenerate, a near-degenerate, or a non-degenerate system refers to this picture and the renormalized energies thereof. In case the electron--phonon
coupling is assumed to be sufficiently strong, such that the polaron shift is much larger than the tunneling-induced broadening $\Gamma$, the system
is effectively non-degenerate, and the canonical rate-equation formalism can be applied. If the polaron shift is of the order $\Gamma$, the electronic
term has to be treated in the singular-coupling limit,\cite{Schultz08c} which will be touched briefly in section~\ref{sec:SCL}. For the setting
with strictly degenerate orbitals that we wish to discuss here, the polaron shift of both levels has to be equal, which limits the possible choices of
the electron--phonon coupling in the above expression to either $\lambda_0$ or $\lambda_z$ being zero.  For strictly degenerate systems, we thus
assume, without loss of generality,
\begin{equation}
H_\text{el--ph} = \lambda \hbar\omega(b^\dagger + b)(n_\uparrow \pm n_\downarrow).
\end{equation}
Choosing the plus sign, we obtain a trivial generalization of the single-mode Anderson--Holstein Hamiltonian.\cite{Koch04} This choice is therefore
termed the ``Anderson--Holstein molecule''. The minus sign, on the contrary, makes the electron-phonon coupling that of an $E\otimes b$ Jahn--Teller
effect.\cite{Bersuker06, Bersuker89}  We call this model the ``Jahn--Teller molecule''. Certain transport properties of Jahn--Teller systems in the
rate-equation regime are analyzed in Ref.~\onlinecite{Schultz08}. By absorbing the sign into orbital specific electron--phonon couplings
$\lambda_\sigma$, the Anderson--Holstein molecule is defined by $\lambda_\uparrow = \lambda_\downarrow$, whereas the Jahn--Teller case is given
when $\lambda_\uparrow = -\lambda_\downarrow$.

\begin{figure*}
  \begin{minipage}[c]{.9\linewidth}
    \includegraphics[width=.9\linewidth]{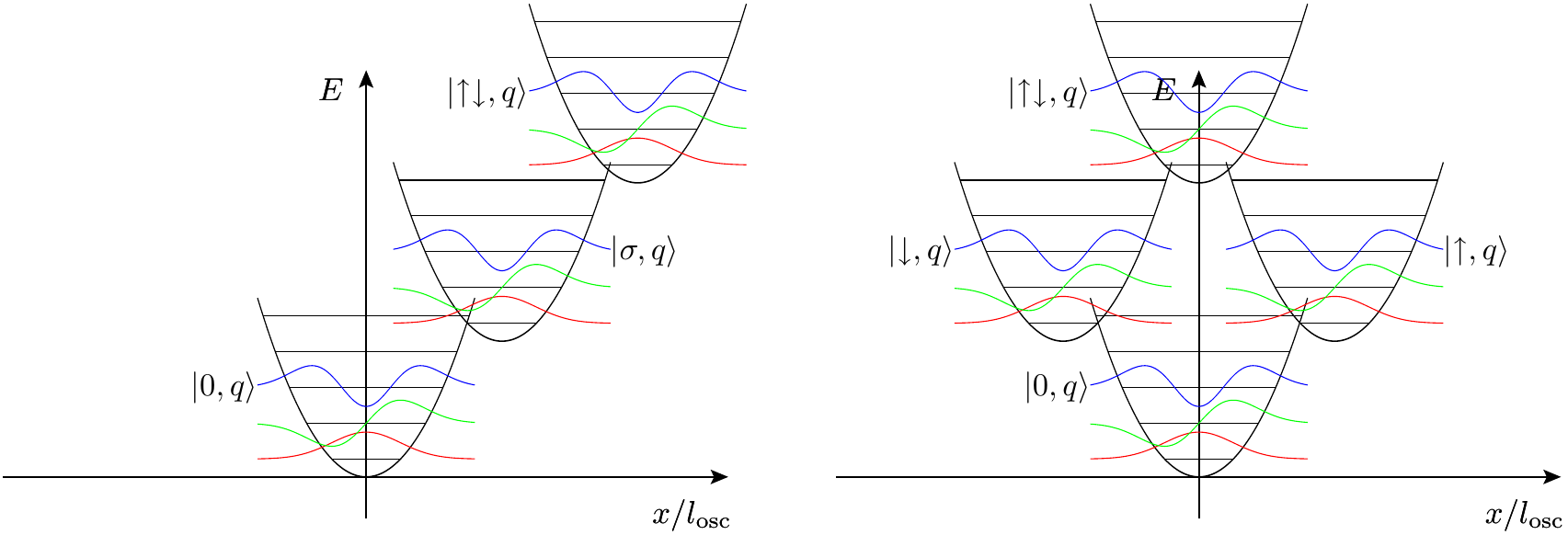}
  \end{minipage}
  \caption{Adiabatic potentials of the Anderson--Holstein molecule (left) and the Jahn--Teller molecule (right). The first few vibronic wavefunctions
  are indicated in order to see the direction dependence of their overlap in the Jahn--Teller case.\label{fig:adiabatic}}
\end{figure*}

The polaron transformation renormalizes not only the single-particle energy $\varepsilon_0 \mapsto \varepsilon_0 - \lambda^2\hbar\omega$ but also the
charging energy $U \mapsto U - 2\sgn(\lambda_\uparrow \lambda_\downarrow) \hbar\omega$.  As in the Jahn--Teller case, both orbitals are shifted to
opposite directions, occupying such a molecule with two electrons will result in a zero net shift of the adiabatic potential, see
Figure~\ref{fig:adiabatic}. The renormalization of the charging energy $U$ will therefore be positive; in contrast to the Anderson--Holstein model,
the Jahn--Teller molecule induces a repulsive interaction between the two electrons in the polaron picture, not an attractive one.\cite{Koch07}

%% >>>

\subsection{Existence of Dark States} %% <<<

As we already know from the treatment of purely electronic structures,\cite{Schultz08c} the master equation can be understood best in a coordinate
system of the degenerate levels, in which one of them is decoupled from the drain electrode. Then a dark state is formed and causes the stationary
current to be strongly suppressed. In the polaron picture, where the tunnel amplitudes are matrix valued, this property is modified.  Consider first
the case without electron--phonon coupling. We apply a unitary transformation in the two-dimensional complex vector space $\mathcal{H}_\text{el}$
spanned by the operators $d_\uparrow$ and~$d_\downarrow$,
\begin{equation}
\begin{pmatrix}
  d_1 \\ d_2
\end{pmatrix}
:=
\begin{pmatrix}
  \cos \theta & e^{\imath \varphi}\sin\theta\\
  -e^{-\imath \varphi}\sin\theta & \cos\theta
\end{pmatrix}
\begin{pmatrix}
  d_\uparrow \\ d_\downarrow
\end{pmatrix}.
\end{equation}
All parts of the Hamiltonian except the tunneling term are invariant under this transformation. The tunneling Hamiltonian becomes
\begin{align}\label{eq:HT_GR}
\tilde{H}_\text{T} = & \sum_{\mb{k}\alpha} c^\dagger_{\mb{k}\alpha}\bigl( (t_{\alpha\uparrow}\cos\theta +
t_{\alpha\downarrow}e^{\imath\varphi}\sin\theta )d_1 \nonumber \\ 
& + (t_{\alpha\downarrow}\cos\theta - t_{\alpha\uparrow}e^{-\imath\varphi}\sin\theta ) d_2\bigr) + \text{h.c.}
\end{align}
By choosing suitable angles $\theta$ and $\varphi$, the second term of one of the above equation vanishes for at least one electrode, allowing the
formation of a dark state: the basis of the coherent current blockade as we have explained in Ref.~\onlinecite{Schultz08c}. Under
certain circumstances, that is for specific choices of the tunnel-couplings, the Hamiltonian can be further simplified.  If the condition
$t_{\alpha\downarrow} cos\theta = t_{\alpha\uparrow}e^{-\imath\varphi}\sin\theta$ can be fulfilled for \textit{every} $\alpha$, the second term in
Eq.~\eqref{eq:HT_GR} will vanish completely and the system will only couple the state $d_1$ to the electrodes, thus reducing it to an effective
single-level system with tunneling matrix elements $\abss{t_\alpha^\text{eff}}^2 = \abss{t_{\alpha \uparrow}}^2 + \abss{t_{\alpha\downarrow}}^2$.
This condition reads
\begin{equation}\label{eq:GR_cond}
\varphi = \arg \frac{t_{\alpha\uparrow}}{t_{\alpha\downarrow}} \quad\text{and}\quad \theta =
\arctan\frac{\abs{t_{\alpha\downarrow}}}{\abs{t_{\alpha\uparrow}}},
\end{equation}
for all $\alpha$. In the following, we shall assume real $t_{\alpha\sigma}$ for simplicity and accordingly set $\varphi = 0$. 

\subsubsection{Anderson--Holstein} %% <<<

In the case of phonons, we have to distinguish the Anderson--Holstein and the Jahn--Teller case. In the polaron picture where the tunnel amplitudes
are  matrix-valued $t_{\alpha\sigma} \mapsto t_{\alpha\sigma} e^{-\lambda_\sigma(b^\dagger - b)}$, the second term of Eq.~\eqref{eq:HT_GR} reads for
electrode $\alpha$
\begin{equation}\label{eq:HT_GR_mol}
c^\dagger_{\mb{k}\alpha} t_{\alpha\downarrow}\cos\theta (e^{-\lambda_{\downarrow}(b^\dagger - b)} - e^{-\lambda_\uparrow(b^\dagger - b)})d_2.
\end{equation}
The electron-phonon coupling of the Anderson--Holstein molecule is characterized by the condition $\lambda_\uparrow = \lambda_{\downarrow}$ thus
rendering the term \eqref{eq:HT_GR_mol} zero. In this model, the decoupling paradigm and the emergence of a dark state works the same way as for
purely electronic levels. In the decoupling regime, such a molecule is equivalent to a single-level molecule. The Jahn--Teller molecule, however, has
$\lambda_\uparrow = -\lambda_{\downarrow}$ making a detailed discussion of the transition rates necessary. 

%% >>>

\subsubsection{Jahn--Teller} %% <<<

Transforming the molecular Hamiltonian into the polaron picture essentially consists of shifting the adiabatic potential of the oscillator by
$\sqrt{2}\lambda_\sigma \ell_\text{osc}$. The Franck--Condon matrix element $M^{qq'}_\sigma = \abss{\fbkt{q}{e^{-\lambda_\sigma(b^\dagger -
b)}}{q'}}^2$ at row $q'$ and column $q$ itself is the overlap of the original vibrational state $\ket{q}$ with the shifted oscillator's state
$\ket{q'}$. The modulus of the matrix elements is independent of the direction of the shift, the sign, however, is not.  A transition between states
that both have an even or odd number of excited quanta still is independent of the direction. A transition from a state with an even number of quanta
to one with an odd number of quanta, however, is sensitive to the direction. In the matrix $e^{\lambda(b^\dagger - b)} - e^{-\lambda(b^\dagger - b)}$,
only the matrix elements belonging to excitations of an even number of oscillator quanta cancel. In the pre-factor of $d_1$, the matrix
$t_{\alpha\uparrow}\cos\theta e^{-\lambda(b^\dagger - b)} + t_{\alpha\downarrow}\sin\theta e^{\lambda(b^\dagger - b)}$, however, there are no
cancellations to be expected. Only in the case of equal coupling, $t_{\alpha\sigma} = 1$ for all $\alpha$ and $\sigma$, the matrix elements for
excitations of an odd number of quanta cancel. In that special case, the Hamiltonian couples transitions with $\abss{q-q'}$ even to $d_2$ and
transitions with $\abss{q-q'}$ odd to $d_1$. Since the both subsets of transistions change the oscillator's energy by either even or odd multiples of
$\hbar\omega$, only, the system decouples into two independent subsystems with disjoint spectra, which in the weak-coupling limit can be treated by
rate equations. As simple and tempting a treatment by rate equations might seem, in this particular case, the rate equation would have two stationary
solutions, such that the asymptotic dynamics would strongly depend on the initial state of the system at time $t=0$.

In the Jahn--Teller configuration, in contrast to the purely electronic model or the Anderson--Holstein molecule, no electronic level can be decoupled
completely from a single electrode by a unitary transformation.  As explicated above, only a certain subset of transitions can be decoupled, the
others remain with finite transition rates. Since this property renders the decoupling paradigm for the Jahn--Teller molecule void at first sight, it
restricts the current-blockade to the voltage regime below the first vibronic sideband, but in return allows to evaluate the model even in the regime
where Eq.~\ref{eq:GR_cond} is satisfied and the Anderson--Holstein model splits up.

%% >>>
%% >>>
%% >>>

\section{Transport Properties} %% <<<

\subsection{General Phenomenology} %% <<<

\begin{figure*}
  \begin{minipage}[c]{\linewidth}
    \includegraphics[width=.45\linewidth]{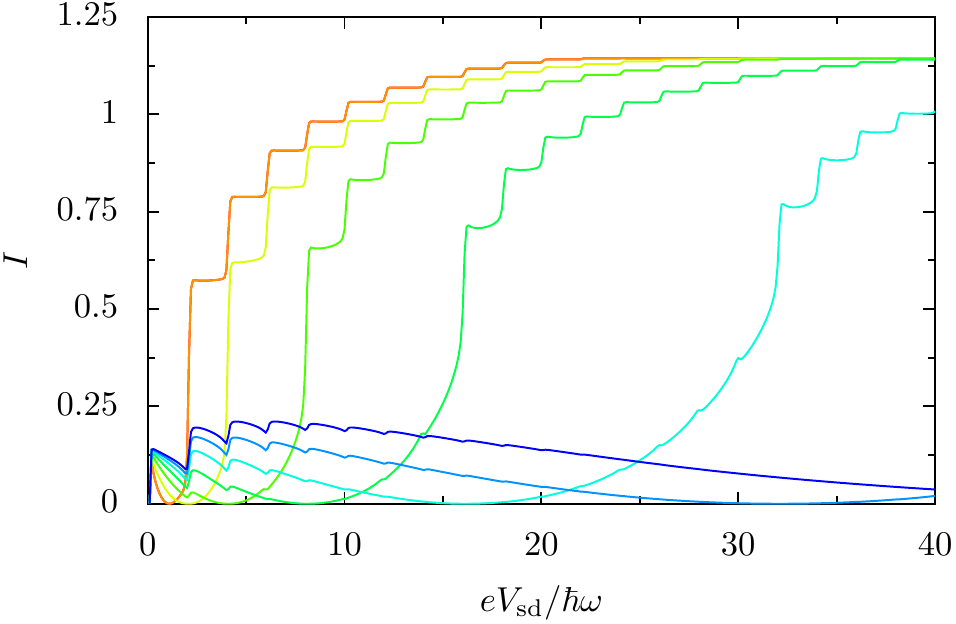}
    \includegraphics[width=.45\linewidth]{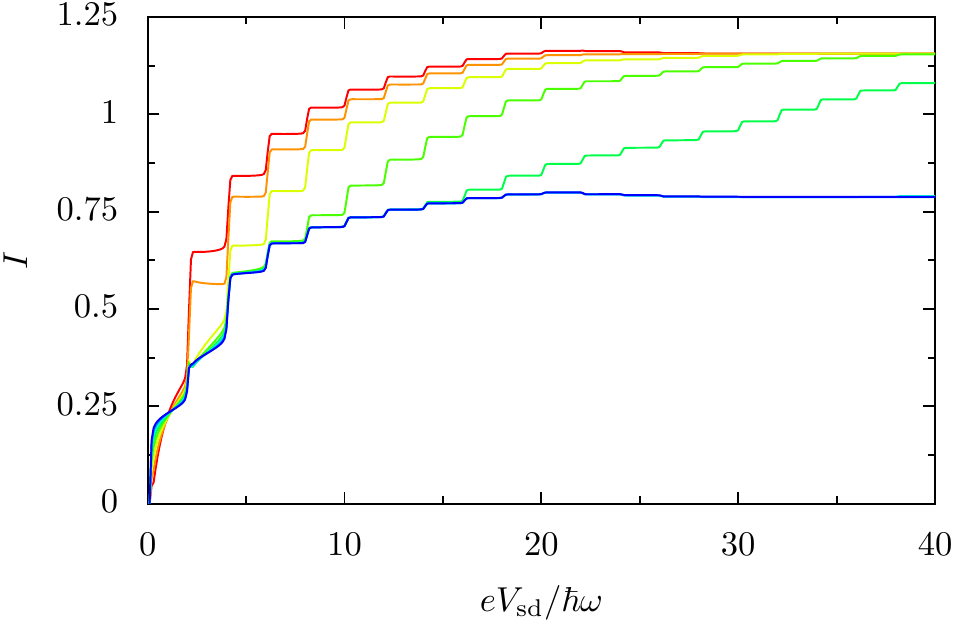}
    \caption{(Color online): Stationary current as a function of the bias $eV_\text{sd}$ for various values of $U/\hbar\omega = 1, 2, 4, 8, 16, 32,
    64, 128$---red to blue curves. The electron--phonon coupling strenght is $\lambda = 1.1$, the tunnel coupling is $\mb{\Gamma} = (1, 1, 1, 2)$. (a)
    Anderson--Holstein molecule. For small $U/\hbar\omega$, the coherent current-blockade is only seen before the first vibronic sideband. For higher
    values of $U/\hbar\omega$, the sidebands appear as a peak-superstructure on the large-scale current-blockade. Since the current-blockade and the
    total inhibition of current flow at $eV_\text{sd} = U/2$ is the dominating feature of this dynamics, the vibronic sidebands become peaks thus
    rendering the appearance of negative differential conductance a generic property of this model. (b) Jahn--Teller molecule. The coherent
    phenomenology in the stationary current is limited to the regime before the first vibronic sideband, as due to the $\sigma_z$ type
    electron--phonon coupling, the decoupling cannot be performed for all vibronic excitations of a given electronic level. \label{fig:Charge}}
  \end{minipage}
\end{figure*}

In the weak-coupling regime $\Gamma \ll \kBT, \hbar\omega$, most transport calculations using rate equations show results, where the Coulomb-blockade
physics of the electronic levels is augmented by a vibronic structure and the appearance of vibronic sidebands in the current--voltage
profile.\cite{Mitra04} Significant diversions from this picture only occur in the case of very weak electron--phonon coupling,\cite{Koch06a} where the
vibronic dynamics heat the molecule, or very strong coupling,\cite{Koch05} where the huge displacement of the oscillators' adiabatic potential causes
an exponential suppression of the low-bias current---the Franck--Condon blockade.

In the presence of two degenerate electronic levels being described by a master equation for the full reduced density matrix, we expect this paradigm
to still hold true. The Coulomb-blockade physics of the electronic structure is enhanced by the coherent current blockade\cite{Schultz08c} due to the
presence of the dark state. Depending on the ratio $U/\hbar\omega$, we shall observe vibronic sidebands in the form of peaks rather than steps
on-top of the profile of the current blockade ($U > \hbar\omega$) or the current blockade modifying the first few vibronic sidebands ($\hbar\omega >
U$). The numerical evaluation of the stationary current for the two generic molecular models shown in Figure~\ref{fig:Charge} corroborates this
reasoning. The Anderson--Holstein molecule complies well with this argument as the formation of the dark state is not influenced by the
vibronic structure. By contrast, the inability of the Jahn--Teller molecule to completely decouple one electronic level shows up in the modification
of the transport properties at the first vibronic sideband only.

The role of the Lamb-shift contributions is in general the same as in the purely electronic case, yielding an additional intra-dot tunneling
Hamiltonian of the form of a pseudo-magnetic field $\mb{B}$, which connects the two degenerate states via virtual intermediate states in the
electronic reservoirs.  The $x$-components of these fields in the pseudo-Bloch equation for $\vec{S}^q$ are, due to the symmetry $\gamma_\alpha^{pq} =
\gamma_\alpha^{qp}$,
\begin{align}\label{eq:pmf_Sq}
B_\alpha^q(\mu_\alpha) = & \sum_p \gamma_\alpha^{pq}\Biggl[\Pint \frac{f(\varepsilon-\mu_\alpha)}{(p-q)\hbar\omega + U + eV_\text{g} - \varepsilon} \D
\varepsilon\nonumber \\
& + \Pint \frac{1-f(\varepsilon-\mu_\alpha)}{(q-p)\hbar\omega + eV_\text{g} - \varepsilon}\,\D \varepsilon\Biggr].
\end{align}
We assume the electronic bands of the leads to be wide enough to ensure $\mathcal{P}\!\!\int\frac{1}{\varepsilon}\D\varepsilon \approx 0$. All
summands vanish simultaneously at $eV_\text{g} - \mu_\alpha = \frac{1}{2}U$, where, similarly to the purely electronic case, the current is completely
blocked: $I = 0$.

Independently of $U$, there is a contribution of the vibronic excitations to the pseudo-magnetic fields, which goes beyond the purely electronic model.
A non-interacting molecule with $U=0$, will still have finite pseudo-magnetic fields as the terms with $p\neq q$ in Equation~(\ref{eq:pmf_Sq}) do not
cancel but yield a contribution that is sharply peaked at the vibronic resonances. This causes slight derivations of the line shape in the
differential conductance from the rate-equation result $\frac{d}{dV} I \propto \frac{d}{dV} f \propto f(1-f)$, the derivative of the Fermi function.

%% >>>

\subsection{Strong Electron--Phonon Coupling} %% <<<

\begin{figure*}
  \begin{minipage}[c]{\linewidth}
    \includegraphics[width=.9\linewidth]{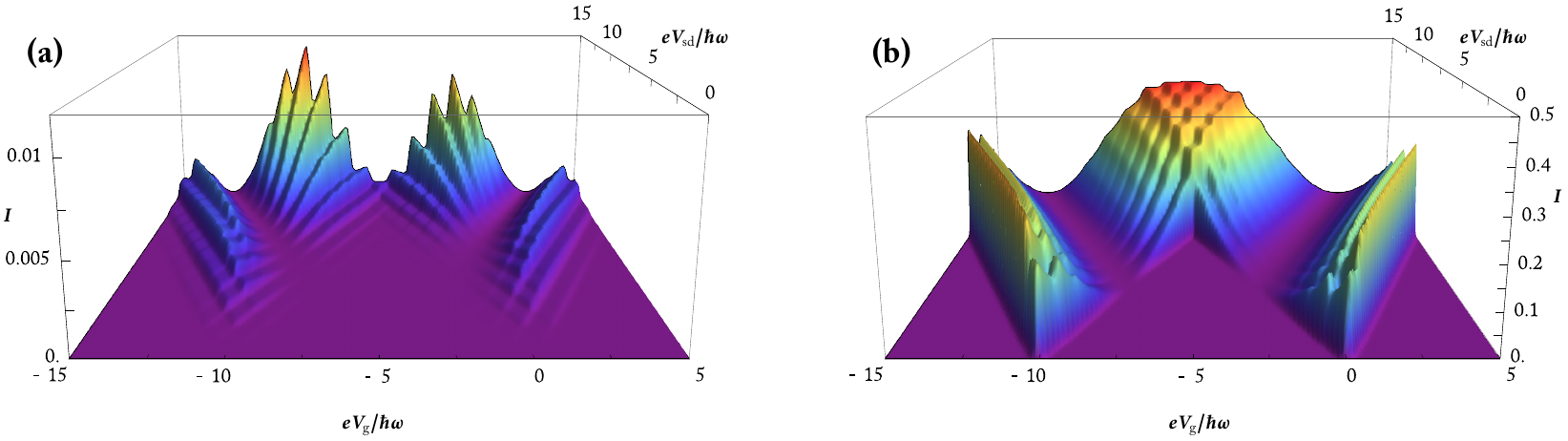}
    \caption{(Color online): Stationary current of the Anderson--Holstein molecule in the Franck--Condon blockade regime for $\lambda = 4$ (a) and for
    intermediate coupling $\lambda = 1.1$ (b). Although the obvious effect of strong electron--phonon coupling is a suppression of the stationary
    current at the Coulomb-blockade line and the double-charging threshold, the current profile develops sharp spikes at larger bias. The parameters are
    $U = 10\hbar\omega$ and $\mb{\Gamma} = (1, 1, 1.5^2, 0)$.\label{fig:FCB_AH}}
     \end{minipage}
\end{figure*}

Similarly to the rate-equation treatment, we expect the molecular models to show significant influence of the vibronic structure on the transport
properties in the regime of strong electron--phonon coupling.\cite{Koch05,Leturcq08} In Figure~\ref{fig:FCB_AH}, we show the stationary current of the
Anderson--Holstein model at intermediate electron--phonon coupling $\lambda = 1.1$ in contrast to strong coupling $\lambda = 4$, where the system is
deep in the Franck--Condon blockade regime.

In Figure~\ref{fig:FCB_AH}~(a), we see the suppression of the low-bias current due to the Franck--Condon blockade.  The additional suppression of the
stationary current at higher bias is not known from the rate-equation model and due to the intrinsic mechanism of the master equation with coherences.
We have shown previously\cite{Schultz08c} and also in this paper that by identifying an exact dark state in the model, we can find the stationary
state of the system. If we can find a state that can be populated easily but leaks only very little probability, this state is still a candidate to
acquire much population in steady-state and apt to dominate the transport physics. Applying this to the large-bias regime of the Anderson--Holstein
molecule in the Franck--Condon blockade yields an explanation for the observed effect. Assume a large charging energy of several $\hbar\omega$ and
bias and gate voltage chosen such that the singly occupied state is aligned slightly above the drain electrode and the doubly occupied state well
below the Fermi level of the source electrode. Then the state $\ket{\sigma,0}$ is almost dark because of the exponentially suppressed Franck--Condon
matrix elements for the transition $\ket{\sigma,0} \mapsto \ket{0,0}$ at the drain. In the conventional rate-equation formalism this would not reduce
the transparency of the device as one could occupy the second level and tunnel off to the drain by converting the large charging energy into vibronic
excitations. In the coherent setting, however, we are allowed to decouple one electronic state from one electrode, and although we have up to now only
decoupled one state from the drain electrode, we could easily decouple one from the source electrode; let this state be $\ket{\bar{\sigma}}$. The
Franck--Condon blockade thus obstructs tunneling to the drain electrode, and the decoupling from the source electrode does not allow to doubly charge
the device, hence reducing the conductance of the device drastically. As before, the Lamb-shift contributions help to regularize this picture, by
transferring the charge from $\ket{\sigma}$ to $\ket{\bar{\sigma}}$, allowing a second electron to occupy the device and converting the charging
energy to vibronic energy in the tunneling to the drain electrode. In this particular setting, the roles of source and drain electrode have been
reversed in the game of decoupling and transferring charge via virtual intermediate state of the reservoirs compared to the low-bias regime.

%% >>>

\subsection{Particular Phenomenology of Jahn--Teller molecules} %% <<<

The electron--phonon coupling of a Jahn--Teller molecule is proportional to $\sigma_z$ in the electronic Hilbert space and thus breaks the rotational
invariance of the molecular Hamiltonian, which we have exploited to understand the steady-state dynamics. We have already shown that this results in
the inability to decouple an electronic level completely from one of the electrodes, thus obstructing the use of generic tunnel couplings. Looking
closer at the quantitative properties of the master equation, the consequences can be seen in a number of effects.

\begin{figure}
  \begin{minipage}[c]{.9\linewidth}
    \includegraphics[width=\linewidth]{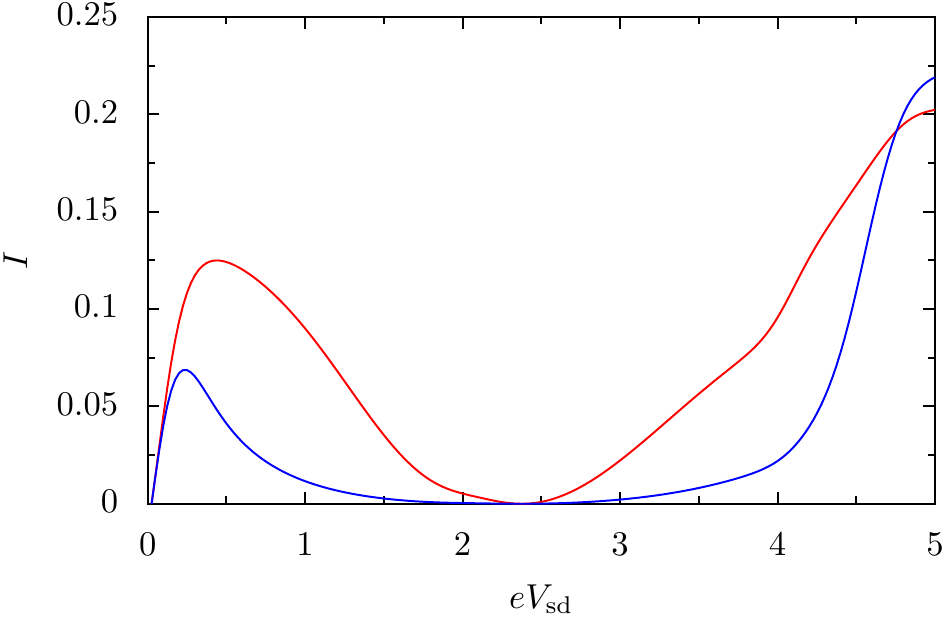}
  \end{minipage}
  \caption{Effects of small pseudo-magnetic field in the $x$-direction. Tunnel coupling $\mb{\Gamma} = (1,1,0,1)$, zero gate voltage, Anderson--Holstein:
  red, Jahn--Teller: blue, $U = 2.35\hbar\omega$\label{fig:Bx_small}}
\end{figure}

The different directions of the shift of the adiabatic potentials implied by the polaron transformation introduce a sign dependence on certain rates
in the master equation. In particular the rates $\gamma_\alpha^{pq}$, will have different signs for $\abss{p-q}$ being even or odd. When we encounter
a situation, where the $x$-component of the pseudo-magnetic fields plays an important role in the dynamics, for example by transferring charge from a
dark state into a conducting one, the structure of the Jahn--Teller molecule's electron--phonon coupling will cause a significant reduction of $B_x$
and thus of the current restoring force. The consequence is, in comparison with the Anderson--Holstein molecule, a profile of the coherent current
blockade, which hardly shows an influence of the Lamb shift at all, having a much higher contrast as is shown in Figure~\ref{fig:Bx_small}.

A second effect of the broken symmetry in the Jahn--Teller molecule is the ability to push the system into the parameter regime, where the
Anderson--Holstein molecule would naturally split into two independent systems, that is, when the ratio $t_{\alpha\uparrow}/t_{\alpha\downarrow}$ is
independent of the lead index $\alpha$ as given by Equation~\ref{eq:GR_cond}. The main difference between the two models is that Jahn--Teller
molecules the pseudo-magnetic fields are non-zero. In this regime, the pseudo-Bloch equations for the model in a voltage regime before the first
vibronic sideband is
\begin{align*}
  \Gamma_\text{S} p_0 - \frac{1}{2}\Gamma_\text{D} p_1 - \frac{1}{2}\Gamma_\text{D} S_z & = B_x S_y\\
  - \Gamma_\text{S} p_0 + \frac{1}{2}\Gamma_\text{D} p_1 + \frac{1}{2}\Gamma_\text{D} S_z & = 0\\
  - \frac{1}{2}\Gamma_\text{D} S_x + B_z S_y & = 0\\
  - \frac{1}{2}\Gamma_\text{D} S_y + (B_x S_y - B_y S_x) & = 0.
\end{align*}
Because $B_x \neq 0$, this equation is \textit{uniquely} solved with $\vec{S} = 0$. A completely vanishing pseudo-spin is not only the statement that
the coherences of the stationary density matrix are evaluated zero but that also the steady-state populations of both electronic levels are equal. Such
a result is unexpected insofar as it is independent of the ratio $\Gamma^\uparrow/\Gamma^\downarrow$, which in a rate-equation treatment would
determine the pseudo-spin's $z$-component. Also, a vanishing pseudo-spin means that the reduced density matrix in the singly-charged sector is
proportional to the unit matrix and invariant under rotations in the electronic Hilbert space.

%% >>>
%% >>>

\section{Modifications of the Basic Model} %% <<<

In this section, we shall discuss several extensions of our basic model of a single-molecule junction with orbital degeneracies, which are related to
experimental issues such as not having exact degeneracy or single-mode reservoirs.

\subsection{Slight Breaking of the Degeneracy}\label{sec:SCL} %% <<<

The requirement of strict degeneracy of the two orbitals is quite unphysical, when one takes into account all the physics of a single-molecule
junction that we have neglected in the abstract model. However, the derivation of a master equation for the reduced density matrix requires strict
degeneracy in order not to sacrifice the density matrix's coherences to the rotating-wave approximation. Although there have been some attempts in the
literature to derive a proper master equation for this case via the, not necessarily positive, Bloch--Redfield equation,\cite{Braun04,Darau08} we
have shown in a previous work\cite{Schultz08c} how such an equation can be derived rigorously by treating the breaking of degeneracy as a
perturbation. 

Assume the degeneracy of the electronic system Hamiltonian $H_\text{S} = (\varepsilon_0 + eV_\text{g})\sum_\sigma d^\dagger_\sigma d_\sigma$, or using the pseudo-spin
notation $H = (\varepsilon_g + eV_\text{g})\Id$, is broken by $H_\Omega:=\frac{1}{2}\Omega \sigma_z$, that is $H_\Omega =
\frac{1}{2}\Omega(d_\uparrow^\dagger d_\uparrow - d_\downarrow^\dagger d_\downarrow)$ in second quantization. Assuming this perturbation to be a slight
breaking of the degeneracy that will not destroy the coherences in the reduced density matrix implies the relation $\Omega \sim \Gamma \ll \kBT,
\hbar\omega$. Following the derivation given in Ref.~\onlinecite{Schultz08c}, we only have to add the Liouvillian $\mathcal{L}_\Omega = [H_\Omega, \cdot]$ to
the master equation found for the degenerate, $\Omega = 0$, model. This term induces a second contribution to the Hamiltonian part of the master
equation besides the Lamb shift; the splitting essentially adds to the $z$-component of the pseudo-magnetic field, but in contrast to the Lamb shift,
$\Omega$ is a free parameter of the theory. Its influence on the steady-state transport properties of our single-molecule devices will be investigated
in the next few sections.

\begin{figure*}
  \begin{minipage}[c]{.9\linewidth}
    \includegraphics[width=\linewidth]{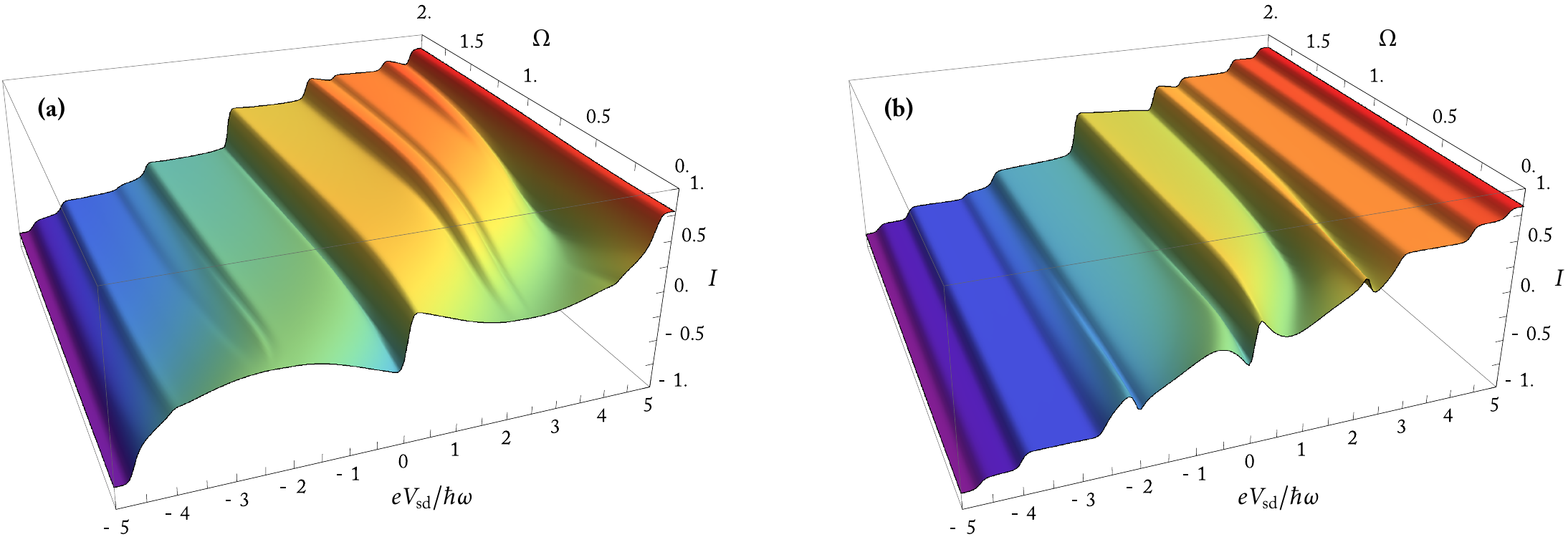}
  \end{minipage}
   \caption{Scan of $\Omega$ at $eV_\text{g} = 0$, $eV_\text{sd} \in [-5\,\hbar\omega, 5\,\hbar\omega]$, $U =
   2.35\,\hbar\omega$, (a) Anderson--Holstein molecule with $\lambda = 1.1$, $\mb{\Gamma} = (1, 1.5^2, 1.5^2, 1)$. The
   splitting is not measured in units of $\hbar\omega$ but, due to the singular-coupling limit $\Omega \sim \Gamma$ in
   units of $\frac{2\pi}{\hbar}\nu_0$. (b) Jahn--Teller molecule for the same parameters.\label{fig:SC-AH-JTE}}
\end{figure*}

\subsubsection{Anderson--Holstein Molecules} %% <<<

As we have shown before, for intermediate electron--phonon coupling, the Anderson--Holstein molecule's behavior is a superposition of the
phenomenology of the electronic levels and the vibronic sideband structure.

In the numerical evaluation of $I(eV_\text{sd})$ as a function of $\Omega$ in Figure~\ref{fig:SC-AH-JTE}~(a), the current suppression is lifted for
larger splitting $\Omega$, and the small vibronic peaks develop into well-defined steps of the current profile, which are typical for the
rate-equation treatment. Due to different tunneling-induced pseudo-magnetic fields $\vec{B}^q$, for each vibronic excitation but constant splitting
$\Omega$, the lifting of the coherent structure and the recovery of the flat profile is different for each sideband. In
Figure~\ref{fig:SC-AH-JTE}~(a), the various vibronic sidebands emerge at different values of $\Omega$.

%% >>>

\subsubsection{Jahn--Teller Molecules} %% <<<

Due to the inability to decouple all vibronic excitations of a single electronic level from the drain electrode in the Jahn--Teller model, the
coherent current blockade being the generic phenomenology of Anderson--Holstein molecules is in general only visible if the zero of the source's
pseudo-magnetic field is at a voltage below the first vibronic sideband. In Figure~\ref{fig:SC-AH-JTE}~(b), we show the $\Omega$-dependent stationary
current at zero gate voltage for intermediate charging energy. As we have claimed, the current blockade is localized close to zero bias and vanishes
quickly as the splitting $\Omega$ is increased. Since the zero of $\vec{B}_\text{S}$, where the current blockade is fully developed, is shifted
towards higher voltages for large charging energy and therefore far beyond the first vibronic sideband, systems with larger Coulomb repulsion will not
show any suppression at all.

\begin{figure*}
  \begin{center}
\begin{minipage}[c]{\linewidth}
  \includegraphics[width = \linewidth]{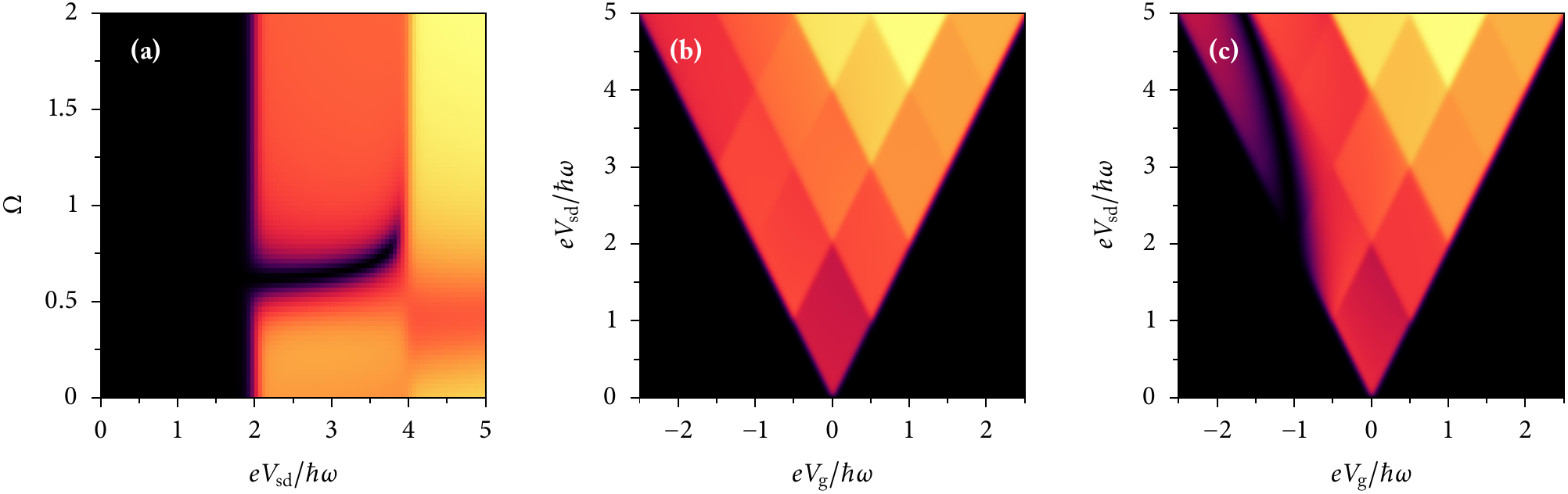}
\end{minipage}
	\caption{(a) Scan of $\Omega$ for the Jahn--Teller molecule at $eV_\text{g} = -\hbar\omega$, $\Gamma = (1.7^2, 1, 1.4^2,
	1)$, $U = 10\,\hbar\omega$, $\lambda = 1.1$, and $\kBT = 0.02\,\hbar\omega$.  (b) stationary current for $\Omega = 0$,
	$eV_\text{sd} \in [0,5\,\hbar\omega]$, $eV_\text{g} \in [-2.5\,\hbar\omega,2.5\,\hbar\omega]$.  (c) same plot, however, with $\Omega =
	0.6$, where similarly to the Anderson--Holstein molecule, a trough in the stationary current due to the
	coherent blockade effect is visible.\label{fig:SC-JTE}}
\end{center}
\end{figure*}

For finite $\Omega$, however, we can shift the zero of $\vec{B}_\text{S}$ so far that it moves below the first vibronic sideband allowing the vibronic
ground state to become dark. We show our findings in Figure~\ref{fig:SC-JTE}, where we plot the current suppression due to \textit{finite} $\Omega$.
Figure~\ref{fig:SC-JTE}~(a) shows the current profile for $eV_\text{g} = -\hbar\omega$ and varying splitting $\Omega$. The dip in the stationary
current for $\Omega \approx 0.6$ is due to the mentioned effect. Diagrams (b) and (c) show a complete current--voltage profile in the $(eV_\text{g},
eV_\text{sd})$-plane. For $\Omega = 0$ (b), the current suppression is absent for all voltages. For $\Omega = 0.6$ (c), on the contrary, there is a
deep trough of suppressed current, slightly reminiscent of the phenomenology of the Anderson--Holstein molecule, but only for the first vibronic
sideband and negative gate voltage.

%% >>>

\subsubsection{General linear electron--phonon coupling} %% <<<

In the introduction, we have considered the general situation of linear electron--phonon couplings. For degenerate electronic systems, these can be
always diagonalized leaving only the identity and the $\sigma_z$ component of the electron--phonon coupling matrix in the electronic Hilbert space
$\mathcal{H}_\text{el}$ $\Lambda = \lambda_0 \Id + \lambda_z \sigma_z$.  Due to the polaron transformation, the system's electronic levels will
acquire an energy difference proportional to $4\lambda_0\lambda_z$.  The strictly degenerate theory, $\Omega = 0$, can only be applied to either the
Anderson--Holstein model, $\lambda_z = 0$, or the Jahn--Teller model, $\lambda_0 = 0$. Using the concept of near-degeneracies, however, we can go
beyond this sharp distinction and consider molecules with both $\lambda_0$ and $\lambda_z$ being non-zero and include the splitting due to the polaron
transformation as a near-degeneracy $\Omega$. Such models are interesting, because the different electron--phonon coupling of the system's levels
amount to different Franck--Condon matrices, for example one level being weakly coupled, the other being in the Franck--Condon blockade regime.

%% >>>
%% >>>

\subsection{Multi-Mode Reservoirs} %% <<<

\begin{figure}
  \begin{minipage}[c]{.9\linewidth}
    \includegraphics[width=\linewidth]{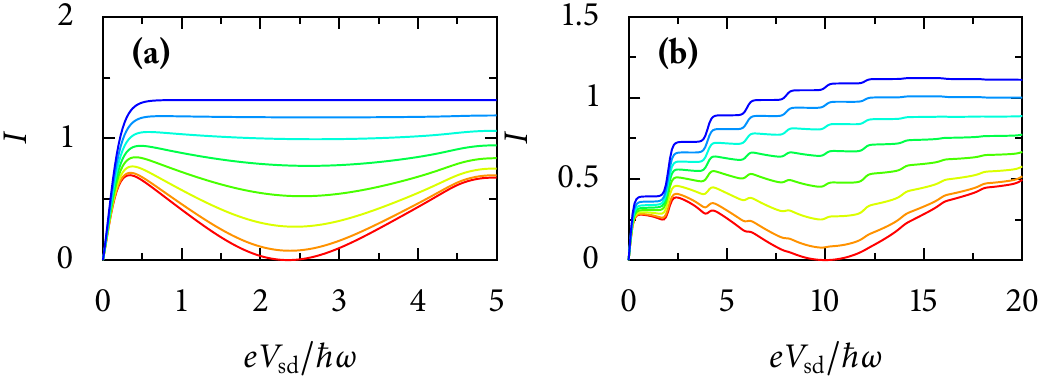}
  \end{minipage}
    \caption{(a) Two-mode reservoir interacting with a two-level quantum dot with $U=2.35\,\hbar\omega$, $\kBT = 0.05\,\hbar\omega$. (b) Two-mode
    reservoir interacting with an Anderson--Holstein molecule.  $U = 10\,\hbar\omega$, $\kBT =0.05\,\hbar\omega$, and $\lambda_\uparrow =
    \lambda_\downarrow = 1.1$. The coupling to the left electrodes are equal and set to unity, and the ones to the right electrode are
    $t_{\text{R}\uparrow,\uparrow} = 1.4$, $t_{\text{R}\uparrow,\downarrow} = t_{\text{R}\downarrow,\uparrow} = 0$. The coupling
    $t_{\text{R}\downarrow, \downarrow}$ is scanned from $0$ to $1.4$ (red to blue curves) in order to increase the eigenvalues of $\Sigma_\text{R}$
    from $(\Sigma^-, \Sigma^+) = (0,1.4)$, where the effect of the two-mode reservoir is absent, linearly to $(\Sigma^-, \Sigma^+) = (1.4,1.4)$, where
    the coherent current blockade is gone.\label{fig:CNT-setup}}
\end{figure}

In a number of recent publications, electronic transport through single molecules is realized by using suspended carbon nanotube quantum
dots.\cite{Sapmaz06,Leturcq08} Following the experimental set-up used in Ref.~\onlinecite{Leturcq08}, a top-gate attached to the suspended nanotube
defines the central piece of it as a quantum dot, the outer pieces as electron supplies directly connected to the metallic source and drain
electrodes, and all three regions being separated through mechanical deformations of the nanotube. The
principal effect of the thus created tunnel barriers to the left and the right of the nanotube quantum dot is that we cannot, in general, assume the
valley quantum number of graphene to be conserved in tunneling anymore. If we denote the valley quantum number of the quantum dot region in accordance
with the notation for degenerate orbitals used throughout this article by $\sigma$ and the valley quantum number of the supplying nanotube electrodes
by $\tau$, the tunnel matrix elements acquire an additional index $\Gamma_{\alpha\tau}^\sigma$. Instead of formerly four, we now have eight tunnel
matrix elements, and it is immediately clear from this number that the paradigm of electronic decoupling at the drain electrode, which we have
established in the discussion of the Anderson--Holstein molecule, is no longer generically true for such systems. The reduction of the master equation
to a rate equation in a suitably chosen basis of the electronic Hilbert space can, however, nonetheless be achieved. Let us consider the
non-interacting retarded self-energy of the tunneling problem with respect to a single electrode, where we assume that the electrodes' distribution function is
diagonal in $\tau$,
\begin{equation}
\Sigma_\alpha \propto \begin{pmatrix}
\sum_\tau \Gamma_{\alpha\tau}^\uparrow & \sum_\tau\gamma_{\alpha,\tau}\\
\sum_\tau\gamma_{\alpha,\tau} & \sum_\tau \Gamma_{\alpha\tau}^\downarrow
\end{pmatrix}.
\end{equation}
This matrix is hermitian and can thus be diagonalized with its off-diagonals being zero. If we choose $\alpha$ to be the drain electrode, the master
equation without the pseudo-magnetic fields is rendered a rate equation.  But now there is no generic dark state and transport does not proceed via
just one electronic levels but including both; however, in a manner known from rate-equation theory. Assume the eigenvalues of $\Sigma_\text{D}$ are
$\Sigma^+ \gg \Sigma^-$. Then the electronic system will accumulate more population in the eigenstate of $\Sigma^-$ and its exit rate towards the
drain electrode will determine the steady-state current. The pseudo-magnetic fields induce an intra-dot tunneling term, which pushes population from
$\ket{\downarrow}$ to $\ket{\uparrow}$ and thus increases the steady-state current when the bias voltage is tuned away from the point $\abss{B} = 0$.
In Fig.~\ref{fig:CNT-setup}, we show numerical results for this model, where we increase the smaller eigenvalue of $\Sigma_\text{D}$ from zero, which
corresponds to the situation found in single-mode reservoirs, to $\Sigma^+$, where the negative differential conductance structure has vanished, for
in that case, both reservoir modes are coupled to the molecule equally well, and the drain's self-energy is proportional to the unit matrix.

The physical significance of this result is the following. The graphs in Fig.~\ref{fig:CNT-setup}~(b) do resemble the current--voltage diagrams
computed in Ref.~\onlinecite{Braun04}. And indeed, the problem of electronic transport with ferromagnetic leads formally resembles our model as there
it is the real electronic spin, which is present both in the electrodes and the quantum dot, but due to the ferromagnetic ordering not conserved in
the tunneling. The primary influence of the presence of more than one electronic mode
in the leads is the lifting of the strict current-blockade and the accompanying softening of the negative differential conductance at the vibronic
sidebands.

%% >>>
%% >>>

\section{Conclusion} %% <<<

The transition from the rate-equation physics of a single spin-degenerate quantum dot to a single-molecule junction is essentially the addition of
inelastic transitions, which first of all cause vibronic sidebands. Only in certain regimes of the model's parameter space, additional vibronic
physics like for instance the Franck--Condon blockade are visible.  In the present article, we have proceeded in a similar manner to extend the
master-equation theory for systems with orbital degeneracies from electronic levels to molecular models, where the main persistent phenomenology is
the addition of vibronic sidebands to the generic current suppression due to the coherent interaction of degenerate or near degenerate electronic
levels.  With two instead of only one level, we have shown that already in the Hamiltonian, the decoupling paradigm, being the essential tool to
understand the transport dynamics, is modified by the matrix structure of the tunnel amplitudes in the polaron picture. We have shown that in the
strictly degenerate case, there are two generic models, which we have termend Anderson--Holstein and Jahn--Teller, with opposite phenomenology. While
the Anderson--Holstein model complies well with the idea that an electron--phonon coupling simply adds the well-known vibronic physics to the
steady-state dynamics of the electronic model, we have shown that the phenomenology of the Jahn--Teller model shows many different transport regimes.
The main reason being the inability to completely decouple an electronic level from one electrode by unitary transformations of the degenerate
orbitals. The decoupling can only be achieved for a certain subset of vibronic excitations, which significantly influences the steady-state properties
of the transport model. 

We have thoroughly discussed the phenomenology of the generic models in various parameter regimes both for the tunneling amplitudes and the
electron--phonon coupling. In the last part of the paper, we have relaxed the requirement of strict degeneracy of the electronic orbitals and allowed
for so-called near-degeneracies, whose influence on the transport properties of purely electronic systems we have already investigated in a previous
publication. We have applied our results to the vibronic models and shown how electronic transport of both models was modified in these regimes. As a
last step, we have turned to the electrodes and discussed the effect of multiple electronic modes in the leads. Our main result is a modification of
the decoupling paradigm used in our theory. Although the coherent current blockade being generic for our models is lifted and, depending on the
parameters, only slightly reduced by the presence of more than one electronic mode in the electrodes, applying unitary transformations in the space of
the degenerate orbitals of the molecule again proves a suitable tool to reduce the master equation to a rate equation for specific voltages and thus
understand its stationary solution intuitively

We are well aware that none of the models, we have discussed so far, would be a complete and quantitative description for any experiment as the models
are too abstract.  However, we have shown that already our models show the generic phenomenology found for more complex systems, for example in
Ref.~\onlinecite{Darau08}, and that we can understand its physics on a very fundamental basis. We have shown how to bring the master equation into a
form where its dynamics can be understood intuitively using basic principles already known from rate-equation theory. With our approach to the theory
of electronic transport through single-molecule devices with orbital degeneracies we therefore provide a basis for future research in the physics of
coherent interactions in sequential tunneling beyond the rate-equation approach. 
%% >>>

\section*{Acknowledgments}
This work was partially supported by SPP 1243 of the Deutsche Forschungsgemeinschaft, during which time I have enjoyed the hospitality of Freie
Universität Berlin, which is courteously acknowledged. Further support has been provided by the Swiss NSF and the NCCR Nanoscience.

\appendix

\section{Derivation of the Master Equation}\label{sec:master-equation} %% <<<

Our aim is the study of the steady-state transport properties of the molecular models in the weak-coupling limit, which amounts to only consider
sequential tunneling processes. To achieve this, we will derive a markovian master equation closely following
Refs.~\cite{SpohnLeb78,Duemcke79,Davies74}. This derivation, whose principal ideas have already been used by us in Ref.~\onlinecite{Schultz08c}
is not contained in the extensive review by Timm,\cite{Timm08} but in our opinion it is the most intuitive and rigorous derivation, because it reduces
the necessary assumptions by a large degree when compared to other approaches. The Hamiltonian of our problem is that of a system--bath interaction
and has the general form
\begin{equation}
H^\xi = H_\text{S} + H_\text{E} + \xi H_\text{S--E}.
\end{equation}
There is a finite-dimensional discrete system $H_\text{S}$, the single molecule, an infinite particle reservoir, the environment $H_\text{E}$, which
in our model are the electronic leads, and a system--bath coupling $H_\text{S--E}$, namely the tunneling Hamiltonian. The parameter $\xi$ in front of
$H_\text{S--E}$ symbolizes the weak-coupling assumption. Later $\xi \to 0$ will be implied and, in order to obtain a non-trivial result, an asymptotic
time scale will be chosen. The quantity of interest is the reduced density matrix of the system $S$, $\rho_\text{S} := \Tr_\text{E}(\rho)$. obtained
by means of a projection operator $\mathcal{P}\rho := \Tr_\text{E}(\rho)\otimes \rho_\text{E} = \rho_\text{S}\otimes\rho_\text{E}$. Also $\mathcal{Q}
:= \Id - \mathcal{P}$. The environment's density matrix will be the equilibrium distribution, which for fermions is just the Fermi-distribution.  The
von Neumann equation $\dot{\rho} = -\imath[H^\xi, \rho] =: -\imath\mathcal{L}^\xi\rho$ is split into an equation for $\mathcal{P}\rho$ and one for
$\mathcal{Q}\rho$ and integrated formally by using the variation-of-constants formula known from the theory of ordinary differential equations, where
$\mathcal{L}_i$ refers to the Liouvillian with respect to $H_i$,
\begin{widetext}
  \begin{equation}
\rho_\text{S}(t) = e^{-\imath\mathcal{L}_\text{S}t}\rho_\text{S}(0)%
- \xi^2\int_0^t e^{-\imath\mathcal{L}_\text{S}(t-s)}\left\{\int_0^s\Tr_\text{E}\left(\mathcal{L}_\text{S--E}e^{-\imath\mathcal{L}^\xi
(s-u)}\mathcal{L}_\text{S--E}\rho_\text{E}\right)\rho_\text{S}(u)\D u\right\} \D s.
\end{equation}
By choosing a factorized density matrix as the initial condition $\rho(0) = \mathcal{P}\rho(0)$, we implement the Born approximation and obtain the
integral equation for the reduced density matrix \textit{alone}.
The substitution $s = u+v$ allows to pull the reduced density matrix out of the inner integral, change the bounds of integration, and move to the
interaction picture with respect to the system Hamiltonian
\begin{equation}\label{eq:Master_Integral_IP}
\rho^\text{I}_\text{S}(t) = \rho^\text{I}_\text{S}(0)%
- \xi^2\int_0^t e^{-\imath\mathcal{L}_\text{S}s}\left\{\int_0^{t-s}
e^{\imath\mathcal{L}_\text{S}v}\Tr_\text{E}\left(\mathcal{L}_\text{S--E}e^{-\imath\mathcal{L}^\xi v}\mathcal{L}_\text{S--E}\rho_\text{E}\right)\D
v\right\} e^{-\imath\mathcal{L}_\text{S} s}\rho^\text{I}_\text{S}(s)\,\D s.
\end{equation}
We aim at deriving a Markovian master equation for which we have to consider a time scale on which all memory effects are absent. This time scale is
defined by the asymptotic time $\tau = \xi^2 t$, which is being held constant in the limiting process $\xi \to 0$, such that
\begin{equation}\label{eq:Master_Integral_IP2}
\rho^\text{I}_\text{S}(\tau) = \rho_\text{S}^\text{I}(0)%
- \xi^2\int_0^\tau e^{\imath\mathcal{L}_\text{S}\frac{\sigma}{\xi^2}}\left\{\int_0^\frac{\tau - \sigma}{\xi^2}
e^{\imath\mathcal{L}_\text{S}v}\Tr_\text{E}\left(\mathcal{L}_\text{S--E}e^{-\imath\mathcal{L}^\xi v}\mathcal{L}_\text{S--E}\rho_\text{E}\right)\D
v\right\} e^{-\imath\mathcal{L}_\text{S} \frac{\sigma}{\xi^2}}\rho^\text{I}_\text{S}(s)\,\D \sigma.
\end{equation}
\end{widetext}
Proceeding with the limit $\xi\to 0$, we find that the operator in curly brackets converges\cite{Davies74} to
\begin{equation}
\mathcal{D} := \int_0^\infty e^{\imath \mathcal{L}_\text{S} v}\Tr_\text{E}\left(\mathcal{L}_\text{S--E}e^{-\imath(\mathcal{L}_\text{S} +
\mathcal{L}_\text{E} )v}\mathcal{L}_\text{S--E}\rho_\text{E} \right)\,\D v,
\end{equation}
An expansion of the Liouvillians in terms of commutators yields the nested commutator structure well known from standard perturbative treatments of
the problem.\cite{Timm08} The action of the two exponential factors $\exp(\pm\imath\mathcal{L}_\text{S}\sigma/\xi^2)$ to the left and right of
$\mathcal{D}$ is to produce the time average $\bar{\mathcal{D}}$ of the operator.\cite{Davies74} The Markovian master equation is thus
\begin{equation}\label{eq:master-equation}
\dot{\rho}_\text{S}(\tau) = - \bar{\mathcal{D}}\rho_\text{S}(\tau),
\end{equation}
whose explicit representation for the models discussed in this paper is given in Appendix~\ref{app:master-equation}. In Refs.~\onlinecite{SpohnLeb78}
and \onlinecite{Davies74}, one finds a very elegant argument, why the time average of $\mathcal{D}$, being an exact result in the limit $\xi\to 0$,
yields the secular or rotating-wave approximation, which is usually applied to decouple the coherences between non-degenerate states from the
respective populations in the reduced density matrix.

We compute the stationary current within the same framework and start from the general definition of the observable. The stationary current through
lead $\alpha$ is
\begin{equation}
\erw{I_\alpha} = \Tr_{\text{S+E}}(I_\alpha \rho_{\text{S+E}}).
\end{equation}
The quantum mechanical current operator for lead $\alpha$ is the time-derivative of the number operator of this electrode,\cite{Wingreen92}
\begin{equation}
I_\alpha = \frac{\imath e}{\hbar}\sum_{\mb{k} \sigma} t_{\alpha\sigma}c^\dagger_{\mb{k}\alpha}d_\sigma - \text{h.c.},
\end{equation}
which is the imaginary part of the operator $\sum_{\mb{k}\sigma}t_{\alpha\sigma}c^\dagger_{\mb{k}\alpha}d_\sigma$. The tunneling Hamiltonian
$H_\text{T}$ is the real part of this operator. We split the expectation value $\erw{I_\alpha}$ into two parts by inserting $\Id = \mathcal{P} + \mathcal{Q}$,
\begin{equation}
\erw{I_\alpha} = \Tr_\text{S+E}(I_\alpha\mathcal{P}\rho_\text{S+E}) + \Tr_\text{S+E}(I_\alpha\mathcal{Q}\rho_\text{S+E}).
\end{equation}
The first term naturally vanishes, because $\mathcal{P}\rho = (\Tr_\text{E}\rho)\otimes \rho_\text{E}$, and the system--bath coupling has zero
expectation value with respect to the equilibrium distribution of the electrodes. It is the projection onto the complement which is the interesting
term. The time evolution of $\mathcal{Q}\rho_\text{S+E}$ is given by\cite{Benatti05}
\begin{align*}
\mathcal{Q}\rho_\text{S+E}(t)  = & e^{-\imath\mathcal{Q}\mathcal{L}_{S+E}\mathcal{Q}t}\mathcal{Q}\rho_\text{S+E}(0)\\%
& - \imath \int_0^t\D s e^{-\imath\mathcal{Q}\mathcal{L}_{S+E}\mathcal{Q}(t-s)}\mathcal{Q}\mathcal{L}_\text{S+E}\mathcal{P}\rho_\text{S+E}(s).
\end{align*}
After integrating the defining equation for the current $I = -\dot{N}$ being the time derivative of the number operator of the respective electrode,
we obtain an integral equation for the latter. We apply the same manipulations as before, using a factorized initial condition for $\rho_\text{S+E}$
that cancels the first term, and changing to the asymptotic Markovian time scale, which shifts the integration boundaries to infinity and takes the
density matrix out of the integral.  The Liouvillian $\mathcal{P}\mathcal{L}_\text{S+E}\mathcal{Q}$ acting on a factorized density matrix is
$\xi\mathcal{L}_\text{S--E}$. The emerging formula is identical to the standard, perturbative derivation, where the factorization of the density
matrix has to be put in by hand but only \textit{after} certain manipulations have been applied to the formula. However, we now know that we can use
the already obtained density matrix from Eq.~\eqref{eq:master-equation} and put it into the formula for the stationary current. An explicit formula in
the basis of the single-molecule device is given in Appendix~\ref{app:master-equation}.

\medskip
%% >>>

\section{Explicit Form of the Master-equation} %% <<<

\subsection{Master equation}\label{app:master-equation} %% <<<
In this section, we give the explicit representation of the  master equation~\eqref{eq:master-equation} for a two-level molecule with arbitrary
Coulomb interaction $U$. The symbols used are defined by $p^q_0  :=  \proj{0,q}$, $p^q_\sigma  :=  \proj{\sigma,q}$, and $p^2_2  :=
\proj{\uparrow\downarrow,q}$, where the first entry specifies the charge state of the molecule and the second the number of excited quanta of the
harmonic oscillator in the polaron picture. The electrons are fermions, which requires to choose a definition of the wavefunction of the doubly
occupied state, $\ket{2}:=\ket{\uparrow\downarrow} = d^\dagger_\downarrow d^\dagger_\uparrow\ket{0}$. The matrix elements
of the molecular terms in the tunneling Hamiltonian therefore differ for having a neutral or a doubly occupied state.
\begin{align*}
\fbkt{\uparrow\downarrow}{t^\ast_{\alpha\downarrow}d^\dagger_{\downarrow}}{\uparrow} & = t^\ast_{\alpha\downarrow} =
\fbkt{\downarrow}{t^\ast_{\alpha\downarrow}d^\dagger_\downarrow}{0}\\
\fbkt{\uparrow\downarrow}{t^\ast_{\alpha\uparrow}d^\dagger_\uparrow}{\downarrow} & = -t^\ast_{\alpha\downarrow} =
-\fbkt{\uparrow}{t^\ast_{\alpha\uparrow}d^\dagger_\uparrow}{0}.
\end{align*}
We also assume only real tunnel matrix elements and a wide-band limit of the electronic reservoirs, that is the density of states is energy
independent and then set to unity for convenience.  We define several short-hands: $t_{\alpha\sigma}^{pq}$ is the tunnel amplitude in the polaron
picture $t_{\alpha\sigma} \fbkt{q}{e^{-\lambda_\sigma(b^\dagger - b)}}{p}$, with $\lambda_\sigma$ incorporating the direction of the adiabatic
potential's shift of the specific model. $\Gamma_{\alpha \sigma}^{pq} := \abs{t_{\alpha \sigma}^{pq}}^2$, $\gamma_\alpha^{pq} :=
t_{\alpha\uparrow}^{qp}t_{\alpha\downarrow}^{\dagger pq} = t_{\alpha\downarrow}^{qp}t_{\alpha\uparrow}^{\dagger pq}$. A missing index on $\Gamma$ or
$\gamma$ indicates that the sum over the respective index is implied. The symbol $\mathcal{P}\!\!\int$ indicates the principal value integral.
\begin{widetext}
  \begin{equation}
\begin{split}
  \dot{p}_0^q = & -2\pi\sum_{\alpha,p} f_\alpha(\varepsilon_1^p-\varepsilon_0^q) \Gamma^{pq}_\alpha p_0^q
  + 2\pi\sum_{\alpha \nu, p}\bigl(1-f_\alpha(\varepsilon_1^p - \varepsilon_0^q)\bigr)\bigl(\Gamma^{pq}_{\alpha\nu} \rho_\nu^p
  + \gamma^{pq}_\alpha\,2\re \rho_{\sigma\bar{\sigma}}^p\bigr)
\end{split}
\end{equation}
\begin{equation}
\begin{split}
  \dot{p}_2^q = & -2\pi\sum_{\alpha, p} (1-f_\alpha(\varepsilon_2^q-\varepsilon_1^p)) \Gamma^{qp}_\alpha p_2^q
  + 2\pi\sum_{\alpha \nu, p} f_\alpha(\varepsilon_2^q - \varepsilon_1^p)\bigl(\Gamma^{qp}_{\alpha\bar{\nu}} \rho_\nu^p
  - \gamma^{qp}_\alpha\,2\re \rho_{\sigma\bar{\sigma}}^p
\end{split}
\end{equation}
\begin{equation}
\begin{split}
  \dot{\rho}_\sigma^q = & -2\pi\sum_{\alpha,p}( f_\alpha(\varepsilon_2^p - \varepsilon_1^q) \Gamma^{pq}_{\alpha\bar{\sigma}} + (1-f_\alpha(\varepsilon_1^q - \varepsilon_0^p)) \Gamma^{qp}_{\alpha\sigma})\rho_\sigma^q
  - 2\pi\sum_{\alpha,p}(-f_\alpha(\varepsilon_2^p - \varepsilon_1^q) \gamma^{pq}_{\alpha} + (1-f_\alpha(\varepsilon_1^q - \varepsilon_0^p)) \gamma^{qp}_{\alpha})\re \rho_{\sigma\bar{\sigma}}^q\\
  & + 2\pi \sum_{\alpha,p}(f_\alpha(\varepsilon_1^q - \varepsilon_0^p) \Gamma^{qp}_{\alpha\sigma} p_0^p + (1-f_\alpha(\varepsilon_2^p - \varepsilon_1^q)) \Gamma^{pq}_{\alpha\bar{\sigma}} p_2^p)
  - 2\mathcal{P}\!\int\D\varepsilon\left\{\frac{f_\alpha(\varepsilon)}{\varepsilon_2^p - \varepsilon_1^q - \varepsilon} \gamma^{pq}_{\alpha} + \frac{1-f_\alpha(\varepsilon)}{\varepsilon_1^q - \varepsilon_0^p - \varepsilon}\gamma^{qp}_{\alpha} \right\}\im \rho_{\sigma\bar{\sigma}}^q
\end{split}
\end{equation}
\begin{equation}
\begin{split}
  \dot{\rho}^q_{\sigma\bar{\sigma}} = & 2\pi\sum_{\alpha,p}\bigl(f_\alpha(\varepsilon_1^q - \varepsilon_0^p)\gamma^{qp}_\alpha p_0^p -
  (1-f_\alpha(\varepsilon_2^p - \varepsilon_1^q)\bigr)\gamma^{pq}_\alpha p_2^p\\
  & - \pi \sum_{\alpha,p}\Bigl[\bigl(-f_\alpha(\varepsilon_2^p - \varepsilon_1^q)\gamma^{pq}_\alpha + (1-f_\alpha(\varepsilon_1^q - \varepsilon_0^p))
  \gamma^{qp}_\alpha\bigr)(\rho_\sigma^q + \rho_{\bar{\sigma}}^q) + \bigl(f_\alpha(\varepsilon_2^p - \varepsilon_1^q)\Gamma^{pq}_\alpha +
  (1-f_\alpha(\varepsilon_1^q - \varepsilon_0^p))\Gamma^{qp}_\alpha\bigr) \rho_{\sigma\bar{\sigma}}^q\Bigr]\\
  & -\imath\mathcal{P}\!\int\!\!\D\varepsilon \left\{\frac{f_\alpha(\varepsilon)}{\varepsilon_2^p - \varepsilon_1^q -
  \varepsilon}\bigl((\Gamma^{pq}_{\alpha\sigma} - \Gamma^{pq}_{\alpha\bar{\sigma}})\rho_{\sigma\bar{\sigma}}^q +
  \gamma_\alpha^{pq}(\rho_{\bar{\sigma}}^q  - \rho_{\sigma}^q)\bigr) + \frac{1-f_\alpha(\varepsilon)}{\varepsilon_1^q - \varepsilon_0^p -
  \varepsilon}\bigl((\Gamma^{qp}_{\alpha\sigma} - \Gamma^{qp}_{\alpha\bar{\sigma}})\rho_{\sigma\bar{\sigma}}^q +
  \gamma_\alpha^{qp}(\rho_{\bar{\sigma}}^q - \rho_\sigma^q) \right\}
\end{split}
\end{equation}
\end{widetext}
Some cosmetic simplifications can be applied by noting that $\Gamma^{pq} = \Gamma^{qp}$ and abbreviating the Fermi factors $f_{\alpha 1}^{pq}:=f(\varepsilon_1^p - \varepsilon_0^q -
\mu_\alpha)$ and $f_{\alpha 2}^{pq} :=f(\varepsilon_2^p - \varepsilon_1^p - \mu_\alpha) = f(\varepsilon_1^q + U - \varepsilon_0^p - \mu_\alpha)$. In
the Anderson--Holstein case, $\lambda_\sigma = \lambda_{\bar{\sigma}}$, also $\gamma^{pq} = \gamma^{qp}$.
The stationary current through lead $\alpha$ can either be computed using the formulae given in section~\ref{sec:master-equation} or derived from the
above equations of motion by noting that this current is given by the $\alpha$-contribution of the expression $\sum_p(\dot{p}_0^p - \dot{p}_2^p)$.
Either way we find
\begin{align}
\frac{1}{2\pi}\erw{I_\alpha} = & \sum_{\sigma, p, q} f_{\alpha 1}^{pq}\Gamma_\alpha^{pq}p_0^p - \sum_{\sigma, p, q}(1-f_{\alpha 2}^{qp})\Gamma_\alpha^{qp}p_2^p \nonumber \\
& - \sum_{\sigma, p, q}\left[(1-f_{\alpha 1}^{pq})\Gamma_{\alpha \sigma}^{pq} - f_{\alpha 2}^{qp}\Gamma_{\alpha\bar{\sigma}}^{qp} \right]\rho_\sigma^p \nonumber \\
& - \sum_{p, q}\left[(1-f_{\alpha 1}^{pq})\gamma_\alpha^{pq} + f_{\alpha 2}^{qp}\gamma_\alpha^{qp} \right]2\re \rho_{\sigma\bar{\sigma}}^p.
\end{align}

%% >>>

\subsection{Pseudo-Bloch representation}\label{app:pseudo-bloch} %% <<<

The master equation \eqref{eq:master-equation} for degenerate two-level systems can be cast into a more intuitive form, since the electronic system
admits the description by a pseudo-spin $\vec{S}$ for the singly charged part of the reduced density matrix. Due to the vibronic structure there is a
pseudo-spin for every excited vibrational state $\ket{\sigma;q}$ there is an additional index $\vec{S}^q$. The components of the pseudo-spin are
naturally defined by
\begin{equation}
\vec{S}^q := \bigl(\re \rho^q_{\uparrow\downarrow}, \im \rho^q_{\uparrow\downarrow}, \rho^q_{\uparrow\uparrow} - \rho^q_{\downarrow\downarrow}\bigr).
\end{equation}
Due to the possibility of changing the charge state, the modulus of this spin is not constant, and the trace of $\rho$, $p^q_1:= \sum_\sigma
\rho^q_{\sigma\sigma}$ as the zeroth component of the Pauli-matrix representation has to also to be included.  Here, we give an explicit
representation of the master equation in terms of the pseudo-spin $\vec{S}^q$ and the populations of the three charging states.
\begin{align}
\dot{\vec{S}}^q = & \pi\sum_{\alpha,p}\Bigl[2f_{\alpha 1}^{qp}p_0^p \vec{n}_\alpha^{qp} + (f_{\alpha 2}^{pq}\vec{n}_\alpha^{pq} - (1-f_{\alpha
1}^{qp})\vec{n}_\alpha^{qp})p_1^q\nonumber \\
& - 2(1-f_{\alpha 2}^{pq})\vec{n}_{\alpha}^{pq}p_2^p \Bigr] - \sum_\alpha\pi\left[f_{\alpha 2}^{pq} + (1-f_{\alpha 1}^{qp}) \right]\Gamma_\alpha^{qp}\vec{S}^q \nonumber \\
&- (\vec{B}_1^{pq} + \vec{B}_2^{qp})\times \vec{S}^q.
\end{align}
The pseudo-magnetization $\vec{n}^{pq}_\alpha$ is defined by the tunnel couplings,
\begin{equation}
\vec{n}^{pq}_\alpha
:=
  \bigl(2\gamma_\alpha^{pq}, 0, \Gamma^{pq}_{\sigma} - \Gamma^{pq}_{\bar{\sigma}}\bigr)
\end{equation}
The pseudo magnetic fields are defined by the principal value terms
\begin{align}
\vec{B}_1^{pq} & := \sum_\alpha\Pint\frac{f_\alpha(\varepsilon)}{\varepsilon_2^p - \varepsilon_1^q - \varepsilon}\D\varepsilon\, \vec{n}_\alpha^{pq},\\
\vec{B}_2^{qp} & := \sum_\alpha\Pint\frac{1-f_\alpha(\varepsilon)}{\varepsilon_1^q - \varepsilon_0^p - \varepsilon}\D\varepsilon\, \vec{n}_\alpha^{qp}.
\end{align}
The equations for the populations are in matrix form
\begin{widetext}
  \begin{equation}\begin{split}
  \frac{d}{dt}\begin{pmatrix}p_0^q\\ p_1^q \\ p_2^q\end{pmatrix} = & \pi \sum_{\alpha p}
    \begin{pmatrix}
      -2f_{\alpha 1}^{pq}\Gamma_\alpha^{pq} & (1-f_{\alpha 1}^{pq})\Gamma_\alpha^{pq} & 0 \\
      2f_{\alpha 1}^{qp}\Gamma_\alpha^{qp} & - f_{\alpha 2}^{pq}\Gamma_\alpha^{pq} - (1-f_{\alpha 1}^{qp})\Gamma_\alpha^{qp} & 2(1-f_{_\alpha 2}^{pq})\Gamma_\alpha^{pq}\\
      0 & f_{\alpha 2}^{qp}\Gamma_\alpha^{qp} & - (1-f_{\alpha 2}^{qp})\Gamma_\alpha^{qp}
    \end{pmatrix}
    \begin{pmatrix} p_0^q \\ p_1^q \\ p_2^q\end{pmatrix}\\
      & + \pi \sum_{\alpha p}\left\{ \begin{pmatrix} 1-f_{\alpha 1}^{pq}\\ 0 \\ 0 \end{pmatrix} \vec{n}_\alpha^{pq} \cdot \vec{S}^p + \begin{pmatrix} 0 \\ f_{\alpha 2}^{qp}\\ 0\end{pmatrix} \vec{n}_\alpha^{qp} \cdot \vec{S}^p + \begin{pmatrix} 0 \\ -(1-f_{\alpha 1}^{qp}) \\ 0 \end{pmatrix} \vec{n}_\alpha^{qp}\cdot\vec{S}^q + \begin{pmatrix} 0 \\ 0 \\ -f_{\alpha 2}^{pq}\end{pmatrix} \vec{n}_\alpha^{pq}\cdot\vec{S}^q\right\}.
\end{split}
\end{equation}
\end{widetext}
By considering the dynamics of the electronic levels only and neglecting any oscillator excitations, the summations over $p$ cancels due to
$\delta_{p,q}$ and we obtain equations similar in form to Ref.~\onlinecite{Braun04}.
%% >>>

%% >>>

%% \bibliography{../../../bibliography/books,../../../bibliography/tunnelling,../../../bibliography/condmat}

\providecommand{\SortNoop}[1]{}

\end{document}